\documentstyle[aps,prl,epsf]{revtex}

\begin{document}

\draft
\twocolumn[\hsize\textwidth\columnwidth\hsize\csname@twocolumnfalse\endcsname

 \title{Instabilities and resistance fluctuations in thin accelerated 
   superconducting rings}
 
\author{Mikko Karttunen$^{1,2}$, K. R. Elder$^{3}$, 
Martin B. Tarlie$^{4}$, and Martin Grant$^{1}$}

\address{$^1$Department of Physics and Centre for the Physics of
Materials, McGill University, 3600 rue University, 
Montr\'eal (Qu\'ebec), Canada H3A 2T8}

\address{$^2$Biophysics and Statistical Mechanics Group, 
Laboratory of Computational Engineering, 
Helsinki University of Technology, 
P.O. Box 9400, FIN-02015 HUT, Finland}

\address{$^3$Department of Physics, Oakland University, Rochester, MI,
48309-4487, U.S.A.}

\address{$^4$James Franck Institute,
University of Chicago, 5640 South Ellis Avenue, 
Chicago, IL 60637, U.S.A.}

\date{\today}

\maketitle

\begin{abstract}
The non-equilibrium properties of a driven quasi-one dimensional 
superconducting ring subjected to a constant electromotive force 
({\it emf}) is studied.  The {\it emf} accelerates the 
superconducting electrons until 
the critical current is reached and a dissipative phase slip occurs that 
lowers the current.  The phase slip phenomena is examined as a function of the 
strength of the {\it emf}, thermal noise, and normal state resistivity.  
Numerical and analytic methods are used to make detailed predictions for 
the magnitude of phase slips and subsequent dissipation.


\end{abstract}

\vskip1pc]
\narrowtext

\section{Introduction}

When driven away from equilibrium, many systems encounter
instabilities leading to new states or phases. Often there exists a
multiplicity of possible states that can be selected near the onset of
the instability.  The selected state may depend on various factors such
as the rate at which the system is driven through the instability,
noise, internal  excitations, different dissipation mechanisms 
and the system size. 

        In this paper the selection of states is studied in  
driven superconducting rings. Many of the phenomena observed
here are not limited to superconducting rings, but appear in
many other physical systems ranging from pattern forming 
systems\cite{kramer85a,benjacob85a,kramer88a,bcaroli91a,hernandezg93a}
to  lasers\cite{torrent88a}. 
The relative simplicity of the superconducting system makes 
it possible to obtain
information about some of the general questions in driven non-linear systems
such as state selection and the effect of dissipation on the state 
selection process itself. 

The mesoscopic nature of the system, i.e., 
the superconducting ring has a finite circumference with a finite
number of accessible states, is fundamental to this problem. First,
it leads to the existence of a finite number of metastable 
current-carrying states which can compete for occupation. It is 
this competition that lies at the heart of the problem. Second, 
care must be taken to distinguish between
voltage-driven and current-driven systems. As shown by 
Tarlie et. al.\cite{tsg94}, for systems that are not in the thermodynamic
limit, i.e. mesoscopic systems, the choice of ensemble is not free. In this
paper we focus on voltage-driven systems as opposed to current-driven systems.

In addition to providing a prototype system to study various aspects
involving driven systems in general, non-equilibrium superconductivity
is of great interest in its own right.
Indeed, the current-induced
transitions in superconducting filaments have been a subject of
intense experimental and theoretical study for almost three decades.
Ref.~\cite{btidecks90a} 
provides a comprehensive review of the field.

We concentrate on the emergence of the dissipative phase-slip
state\cite{rieger72a,skocpol74a,kramer77a,kramer84a,tarlie98b} in
voltage-driven mesoscopic systems.  When a superconductor (below $T_c$) is 
driven by a voltage-source, the supercurrent increases until it reaches a 
critical value at which point the system becomes unstable.  
Several interesting phenomena may then occur: the system will enter the dissipative
phase-slip state, Joule heating can take place, mode locking, as well
as other phenomena.   Here, the focus is on the onset of
the instability and its effect on the dynamics of the superconducting
state.

The transitions between the current-carrying states can take place via
two fundamentally different routes: {\it i)} by a nucleation process
involving thermal fluctuations and an activation energy
barrier, or {\it ii)} the system may be driven to an instability by an
external driving force.  In the context of nucleation and
metastability, the decay of persistent currents in thin
superconductors is an old and extensively studied
problem\cite{little67a,langer67c,mccumber68a,mccumber70a}.  However,
the latter\cite{tarlie98b} involves a decay from a point of
instability, and it is relatively poorly understood.  One of
the major difficulties is this: whereas in the case of nucleation the
decay is from a metastable state involving thermal activation and a
saddle point, in the latter case the external force drives the system
to a point of instability where there is no energy barrier left, i.e.,
the energy landscape looks locally flat.  In this instance the decay 
and the final state depend on various factors, such as how fast the system was
driven, the relative strength of fluctuations, internal excitations,
and so on. This makes precise theoretical formulation of the problem
difficult, since it is not possible to use the free energy formulation 
as in the case of metastability\cite{mckane01a}.

\section{The system}

\begin{figure}[!]
\centerline{
\epsfxsize=\columnwidth 
\epsfbox{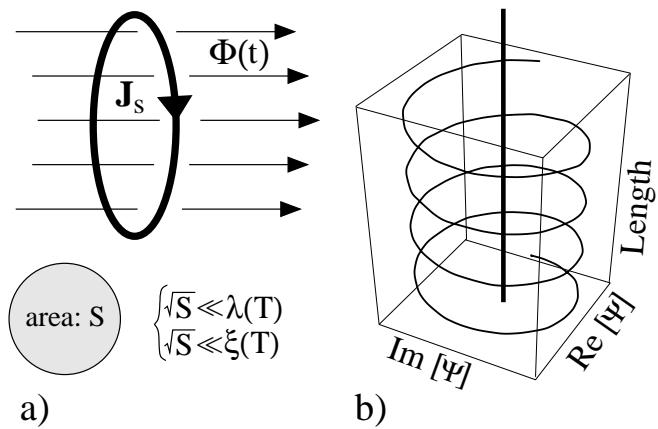}
}
\vspace*{0.2cm}
\caption[Illustration of a voltage driven superconducting ring]
{a) Illustration of a voltage driven superconducting ring.
The magnetic flux is due to an infinitely 
long solenoid passing through the center of the loop. 
b) Illustration of the current carrying states as uniformly twisted plane waves. 
At a phase slip the amplitude of the helix approaches locally zero.
The line through the center of the helix represents the superconducting wire.
}
\label{fig:system}
\end{figure}

The physical system considered is a quasi-one-dimensional
superconducting ring of finite circumference, 
i.e., the radius of the cross-section area ($S$) of the superconducting filament 
is much smaller than coherence length ($\xi$) and  
magnetic penetration length ($\lambda$): $\sqrt{S} \ll \xi(T)$
and $\sqrt{S} \ll \lambda(T)$, respectively, see Fig.~\ref{fig:system}a. 
When the ring is placed in a time-dependent magnetic field, by 
Faraday's law of induction, an emf is induced in the ring. 
From London's equation ($\vec{E}(\vec{r}\,)
=\partial_t[(4 \pi \lambda^2/c^2)
\vec{J_s}(\vec{r},t)]$) this leads to a 
current that increases in time. (Here $\vec{E}$ is the electric
field, $\vec{J}_s$ the supercurrent density, and $c$ the speed of light.)
The time-dependent increase in the current cannot continue indefinitely, 
and eventually the current will reach a critical value, at which point the
system becomes unstable and a dissipative phase slip will occur, resulting 
in a reduction of the current\cite{tarlie98b} by a discrete amount. 

It is important to reemphasize that the system dynamics in the case under
study here, viz. the decay of the system from a point of {\it instability},
is very different from the historically well-studied problem of the decay of
the system from a point of {\it metastability}. The picture of
the system hopping from one local minimum to the next 
no longer applies. Rather, the picture now is one where the system 
is initially in a locally stable state, but as a consequence of the voltage source,
the energy landscape evolves in such a way that as the critical current is reached,
the system finds itself at the top of a hill. When this situation is encountered, 
it is possible that there exist a variety of different valleys for the system
to flow into, each valley leading to a locally stable state. In this picture, 
each of these locally stable states compete for occupation.

To examine this phenomena the Ginzburg-Landau theory of 
dirty superconductors will be considered. 
The Ginzburg-Landau free energy functional can be 
written, 
\begin{eqnarray}
{\cal F} \left[\Psi,\vec{A} \right]  =  \int d \vec{x} \left\{ \frac{\hbar^2}{2m_e}
\left| \left( \nabla - i \frac{2e}{\hbar c} \vec{A} \right) \Psi
\right|^2  + \right. \nonumber \\ 
a \left| \Psi \right|^2 
\left. + \frac{b}{2} \left| \Psi \right|^4  \right\} 
+ (8 \pi)^{-1} \int d \vec{x} (\nabla \times \vec{A})^2,
\label{eq:glfree}
\end{eqnarray}
where $\vec{A}$ is the vector potential,
$\Psi$ is the complex valued order parameter, 
$e$ is the electron charge, $m_e$ is the electron mass,
$c$ is the speed of light, $\hbar$ is Planck's constant,
and $a$ and $b$ are the expansion coefficients.
Since the current is induced in the loop by a time varying magnetic flux,
the effect of the induced electromotive force (emf) must be included 
in the GL description. 
By Faraday's law of induction, the electrons in the loop are subjected 
to an  emf
\begin{equation}
{\cal E} = -\frac{d \Phi(t)}{d t} = \oint \vec{E} \cdot d \vec{l},
\label{eq:faraday}
\end{equation}
where $\cal{E}$ is the induced emf and $\Phi(t)$ is magnetic flux
through the loop.  The magnetic flux and the
magnetic field are related by 
\begin{equation}
\Phi = \int\limits_{area} \vec{B} \cdot d \vec{S}_l,
\label{eq:mflux}
\end{equation}
where $\vec{S}_l$ is the area of the loop. 
Eqs.~(\ref{eq:faraday}) and (\ref{eq:mflux}) can 
be combined to obtain a relationship between 
the the vector potential and the electric field, i.e., 
if $E_x$ is used to denote the tangential component of the field then 
$A_x=-E_x ct,$ where $A_x$ is the tangential component of the vector  
potential.  This in turn gives $A_x = -{\cal E} ct/L$, where 
$L$ is the length of the wire.

The one-dimensional nature of the problem allows 
several simplifications. First, 
since the wire is narrow the magnetic field generated 
by the supercurrent does not significantly influence the order 
parameter. This allows one to treat the 
vector potential $A_x$ as a parameter instead of as a dynamical variable. 
In addition, since the magnetic field energy due to the supercurrent is much
smaller than the energy associated with the order parameter, the
magnetic field term can be dropped from the free energy\cite{langer67c}.
Finally, since the radius of the wire is less than $\xi$ the order parameter 
is only a function of the tangential direction ($x$). 
The geometry of the wire implies periodic boundary conditions, i.e., 
$\Psi(x)=\Psi(x+L)$.

For further analysis and computational efficiency it is 
convenient to rewrite the equation in dimensionless form
using the following transformations:
\begin{eqnarray}
\Psi'  & \equiv & \left(b/ |a|\right)^{1/2} \Psi \nonumber \\
x' & \equiv & x/\xi \nonumber \\
A'  & \equiv & 2e \xi A/ \hbar c  \nonumber \\
v_{ec}'& \equiv &  v_{ec}2e\tau_{GL}/\hbar
\label{eq:trans}
\end{eqnarray}
where $\xi^2= \hbar^2/(2m_e |a|)$ and it is implicitly assumed that 
the temperature is below the superconducting transition (i.e., 
$a < 0$).  $\tau_{GL}$ is the Ginzburg-Landau time 
defined as 
\begin{equation}
\tau_{GL}=\frac{\pi \hbar}{8 k_B (T_c-T)},
\label{eq:gltime}
\end{equation}
and it is the natural measure for time, i.e., $t\rightarrow t/\tau_{GL}$.
In the following, we will work in dimensionless units, i.e., we perform
the transformations as defined above and drop out the primes for convenience.

The last transformation in Eq.~(\ref{eq:trans}) involves $v_{ec}$, 
the electrochemical potential generated by the normal current, which 
will be formally introduced in the next section where the GL 
theory is extended to include normal (Ohmic) current generation. 
In addition, following the scalings in Eqs.~(\ref{eq:trans}), it is 
natural to measure the length in units of the coherence length as
$\ell=L/\xi$.  Then the rescaled  boundary condition reads
$\Psi(x)=\Psi(x+\ell)$,
and 
the dimensionless free energy 
becomes
\begin{equation}
{\cal F}=\int_{-\ell/2}^{\ell/2} dx  \left\{\left| \left(\partial_{x} 
-i A_{x} \right) \Psi \right|^2 - |\Psi|^2 
+\frac{1}{2}|\Psi|^4 \right\}.
\end{equation}

        To describe the dynamics of the superconducting 
condensate, relaxational dynamics are assumed leading to 
the standard stochastic time-dependent
Ginzburg-Landau  (STDGL) equation of motion, i.e., 
\begin{equation}
\frac{\partial \Psi}{\partial t}= 
-\frac{\delta {\cal F}}{\delta \Psi^*} + \eta,
\label{eq:langevin}
\end{equation}
where  $\eta \equiv \eta(x,t)$ is an uncorrelated Gaussian 
noise source with correlations
\begin{eqnarray}{}
\langle \eta(x,t) \rangle &=&0 \nonumber \\
\langle \eta^*(x,t)\eta(x',t') \rangle &=&
2D \delta(x-x') \delta(t-t'). \nonumber
\end{eqnarray}
The angular brackets denote an average, and 
$D$ is the intensity of the noise 
determined by the fluctuation-dissipation\cite{mccumber70a} theorem as 
\begin{equation}
D=\frac{2 \pi k_BT}{SH_c^2\xi},
\label{eq:thermnoise}
\end{equation}
where $H_c$ is the critical field, and  $H_c^2 \propto (1-t)^2$,
$\xi(T) \propto (1-t)^{-1/2}$ and  $t=T/T_c$
\cite{btinkham96a}.

To make the model numerically more tractable, it is 
convenient to make the 
transformation\cite{mccumber70a,byers61a}
$\Psi \rightarrow \Psi e^{iq (t)x}$, where 
\begin{equation}
q(t)=A_x =  \omega \ell^{-1} t,
\label{eq:qt}
\end{equation}
where $\omega=\tau_{GL}2e{\cal E}/\hbar$.
This transformation twists, or winds,
the order parameter along the wire. The effect of the transformation is to map 
the current carrying states to twisted plane waves
as illustrated in Fig.~\ref{fig:system}b.
After the transformation, the periodic boundary condition becomes
\begin{equation}
\Psi(\ell +x,t)=\Psi(x,t)e^{iq(t)\ell },
\end{equation}
and the equation of motion obtained from Eq.~(\ref{eq:langevin}) reads as
\begin{equation}
\frac{\partial \Psi}{\partial t} = 
\frac{\partial^2 \Psi}{\partial x^2} + \Psi -\Psi \left| \Psi \right|^2
+ i \ell^{-1} \omega x \Psi + \eta.
\label{eq:transtdgl} 
\end{equation}
This formulation neglects the electrochemical potential
due to normal current generation at a phase slip center. 
Its inclusion is discussed  next. 

\subsection{Electrochemical potential}

Eq.~(\ref{eq:transtdgl}) would be a sufficient description if the 
generation of a normal current at a phase slip could be neglected.
This approximation is valid when the normal state resistivity 
is negligible\cite{tarlie98b,mccumber70a}.  However, the Ginzburg-Landau 
free energy is only valid for `dirty' superconductors in which the normal 
state resistivity is appreciable even at low temperatures.    One
aim of the current study is to examine the effect of the resistive normal 
current to the process.  To facilitate this goal 
the equation of motion (i.e., Eq.~(\ref{eq:transtdgl})) must be 
generalized to include the creation of electrochemical potential gradients 
at phase slip locations.

A phase slip occurs when the system locally loses superconductivity and
becomes a normal Ohmic conductor. As discussed above, below $T_c$ the system
retains the fully superconducting state after making a transition to 
a state of lower current. 
An important question is the effect of the generation of 
normal current on the dynamics and the state selection problem. 

To account for the generation of normal current, 
the time derivative in the STDGL equation of motion 
must be replaced by $\partial/\partial t + iv_{ec}$, 
where $v_{ec}\equiv v_{ec}(x,t)$ is the electrochemical potential generated 
by the normal current\cite{mccumber70a,stephen64a,anderson65a,schmid66a,abrahams66a}.
With that substitution, the dimensionless equation of motion becomes
\begin{equation}
\left(\frac{\partial}{\partial t} + iv_{ec}\right)\Psi =  \partial_x^2 \Psi 
+ \Psi -\Psi \left| \Psi \right|^2  + i \frac{\omega x}{\ell} \Psi + \eta.
\label{eq:model}
\end{equation}

Physically, the appearance of the electrochemical potential is due to local
charge imbalance in a superconductor. 
Gorkov\cite{gorkov58a} was the first to point out that, in a superconductor
the Fermi level, and thus the electrochemical potential is a local
time-dependent variable related to the coherence of the 
superconducting state.
Qualitatively, if the local charge balance is disturbed, the Fermi level 
experiences a local time-dependent perturbation. This in turn affects
the local energy gap. Gorkov showed that gauge invariance is preserved,
if the order parameter depends on time as $\exp(-2i\mu_F t/\hbar)$ 
where $\mu_F$ is the Fermi energy.  This leads to the
second term on the left hand side in Eq.~(\ref{eq:model}).

        The electrochemical potential can be determined by combining 
charge conservation and Ohm's law in the following manner. 
Charge conservation implies that,
$\partial_x(J_n+J_s) = 0,$ 
where $J_n$ is the normal current and $J_s$ is the supercurrent\cite{superC}.
From Ohm's law, i.e., $\partial_x v_{ec} = -\alpha\, J_n$, this 
can be written
\begin{equation}
\frac{\partial^2 v_{ec}}{\partial x^2} = \alpha \frac{\partial J_s
} {\partial x},
\label{eq:echem}
\end{equation}
where $\alpha$ is a dimensionless Ohmic resistivity
and can be written
\begin{equation}
\alpha = \rho_n/\rho_o
\label{eq:alpha}
\end{equation}
where $\rho_n$ is the normal state resistivity, 
\begin{equation}
\rho_o = \frac{k_B T_c(1-t)\hbar}{\pi \xi(T)^2e^2H_c(T)^2},
\label{eq:resiso}
\end{equation}
$t=T/T_c$ and $H_c(T)$ is the critical field.
For a dirty superconductor\cite{btinkham96a} this can be written
\begin{equation}
\rho_o = 0.1455\times \frac{\hbar \mu_ok_BT_c}{\xi_o l_F e^2H_c^2(0)} 
\label{eq:resisod}
\end{equation}
where $l_F$ is the mean free path length.

\section{Linear stability analysis}
\label{sec:lsa}

The aim of the linear stability analysis is to gain insight into 
the stability of the current carrying state against small perturbations, 
how the perturbations grow or decay in time, and how different 
modes are selected.  In general, when the total current exceeds
the critical supercurrent an Eckhaus instability occurs. 
The Eckhaus instability is a longitudinal secondary instability 
that appears in many systems exhibiting spatially periodic 
patterns\cite{beckhaus65a,dominquez86a}

To study the Eckhaus instability in superconducting rings the 
order parameter is linearized around a current carrying state 
by setting $\Psi(x,t)=\Psi_0 +\delta \Psi(x,t)$, where 
$\Psi_0=\sqrt{1-q^2}\,e^{iqx}$ and $q=(\omega/\ell)t$. 
$\Psi_0$ is a current carrying (or uniformly twisted plane wave) 
state that is a solution of Eq.~(\ref{eq:model}) in the 
limit $\omega/\ell \ll (1-q_c^2)^2/q_c \approx 0.77$, where 
$q_c = 1/\sqrt{3}$.  This limit is satisfied for the range 
of $\omega/\ell$'s considered in this paper (i.e., 
$2 \times 10^{-6} < \omega/\ell < 2 \times 10^{-3}$).
Since the system possesses 
translational invariance and admits plane wave solutions, 
the perturbation is given in terms of its Fourier expansion, i.e.,
$$
\delta \Psi(x,t)=\sum_{n}\left( a_{k_n}(t) 
e^{ik_{n}x}+a_{-k_n}(t)e^{-ik_nx}\right)e^{iqx},
$$
where $a_{k_n}(t)$ is the amplitude of mode $n$ associated with wavevector 
$k_n =2\pi n/\ell$.  Substituting into Eq.~(\ref{eq:model}), using
Eqs.~(\ref{eq:qt}) and (\ref{eq:echem}) to solve for 
$v_{ec}$ and linearizing in $\delta \Psi$ gives an equation of 
motion for $\delta \Psi$ or in Fourier space for $a_{k_n}$.  Setting 
$a_{k_n}(t) = a_{k_n}e^{\lambda(q,\alpha)t}$
leads to an eigenvalue equation which can be solved to give
\begin{eqnarray}
\lambda^{\pm}_n(q,\alpha) 
=&-&(1-{q}^{2})(1+\alpha/2)-k_n^2 \nonumber \\
&\pm& \Big[(1-{q}^{2})^2\left(1-\alpha+\alpha^2/4\right) \nonumber \\
&\,& +4q^2\left(k_n^2+\alpha (1-{q}^{2})\right)\Big]^{1/2}.
\label{eq:eigen}
\end{eqnarray}

\begin{figure}[!]
\centerline{
\epsfxsize=\columnwidth
\epsfbox{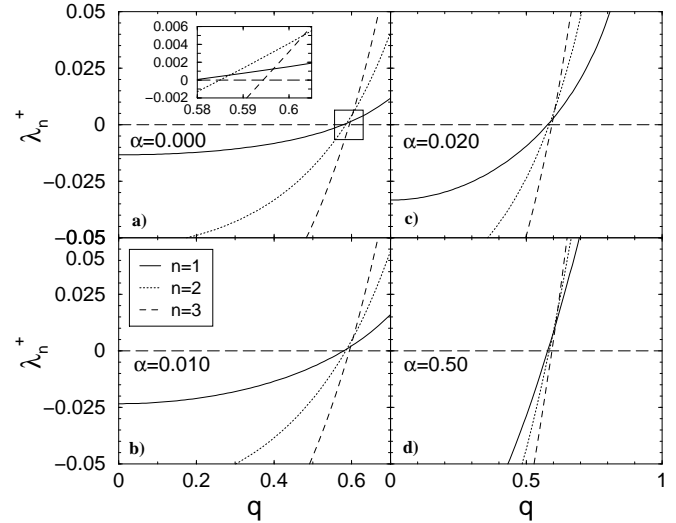}
}
\caption[The eigenvalue spectrum from linear analysis]
{The eigenvalues as a function $q(t)=\omega t / \ell$ for
different $\alpha$'s. The solid line is the
$n=1$ mode, dotted line $n=2$, and dashed line $n=3$ mode.
For small $q$ all the modes are stable, i.e., $\lambda_n^+ <0$,
independent of the value of $\alpha$. Increasing the dissipation 
($\alpha$) increases the growth rate of all phase slips modes.
The caption in a) shows the boxed area.}
\label{fig:linear}
\end{figure}

When $\lambda_n^{\pm}$ is negative the corresponding mode is
stable, fluctuations decay back to zero, and the superconducting 
state persists.  When $\lambda_n^{\pm}$ is positive the 
current carrying states are unstable with respect to fluctuations 
of a finite wavevector $k_n$.  For the following discussion 
$\lambda_n^{-}$ can be neglected as it is negative definite.

        In Fig.~\ref{fig:linear}, $\lambda^+$ is shown for the 
first three modes as a function of $q$ for several different 
values of $\alpha$.  For small $q$ all the modes are 
stable, i.e., $\lambda_n^+ <0$.  The inset in Fig.~\ref{fig:linear}a 
shows that the modes become unstable sequentially; the lowest mode first, 
then the mode $n=2$, and so on.  The time ($t_n$) at which a given mode 
becomes linearly unstable is determined by the condition 
$\lambda_n(t_n) = 0$ which gives
\begin{equation}
t_{n1} = \frac{\ell}{\omega} \sqrt{1/3+k_n^2/6}.
\label{eq:tn}
\end{equation}
For a wire of infinite length this time corresponds to the time at 
which the current reaches the critical value, i.e., $q_n = (w/\ell) t_{n1} 
= \sqrt{1/3+k_n^2/6} \rightarrow \sqrt{1/3}$ and 
$J_c = q_c(1-q_c^2) = 2/\sqrt{27}$.  While Eq.~(\ref{eq:tn}) 
implies that single phase slip (i.e., $n=1$) processes will dominate, 
this effect is offset by the rate of increase of $\lambda^+_n$, 
i.e., $\partial \lambda^+_n/\partial q$ is an increasing function 
of $n$.  This can be seen in the small $\alpha$ limit, i.e., 
\begin{eqnarray}
\left.\frac{\partial  \lambda^+_n}{\partial q}\right|_{q_n} 
&=& \frac{2\sqrt{12+6k_n^2}}{4+5k_n^2}\Big(3k_n^2 +\nonumber \\
&\,&\left.+\frac{(k_n^2+2)(4-k_n^2)}{4+5k_n^2}\alpha + {\cal O}(\alpha^2) \cdots \right),
\end{eqnarray}
or for simplicity in the small $k_n$ limit
\begin{equation}
\left.\frac{\partial  \lambda^+_n}{\partial q}\right|_{q_n}
= \sqrt{3}\,(3k_n^2+2\alpha + \cdots).
\end{equation}
Thus the rate of increase of the positive eigenvalue 
increases with $n$.  The situation is somewhat analogous to the 
classic tortoise/hare race if only two modes are 
considered (say $n=1$ and $2$).  In this case the tortoise 
($n=1$) begins the race first since $t_1 < t_2$, but the 
hare accelerates faster since 
$\partial \lambda^+_1/\partial q|_{q_1} < \partial \lambda^+_2/\partial 
q|_{q_2}$.  To first order in $\alpha$, the effect of dissipation is 
to increase the rate of acceleration of both the tortoise and 
hare equally.  Since the tortoise begins the race first this tends 
to favor the tortoise winning the race.  In terms of mode analysis, 
increasing the dissipation (i.e., $\alpha$) increases the probability of 
a single phase slip ($n=1$) occurring over a double phase slip ($n=2$).

        The linear predictions can be used 
to estimate the relative probabilities of a phase slip of order 
$n$ occurring.  In the linear prediction the equal time correlation function 
for the $n^{th}$ mode is
\begin{equation}
\langle |a_n(t)|^2 \rangle = \frac{2D}{\ell} e^{2\sigma(t,\alpha)}
\int_0^t dt' e^{-2\sigma(t',\alpha)}
\label{eq:an2}
\end{equation}
where
\begin{equation}
\sigma(t,\alpha) \equiv \int_0^t dt' \lambda^+_n(t',\alpha).
\end{equation}
Following the instability Eq.~(\ref{eq:an2}) describes the evolution 
of the $n^{th}$ mode from the initial current carrying state 
described by $\Psi^i_n = \sqrt{1-q^2}{\rm exp}[iqx]$ 
to the new current carrying state 
described by $\Psi_n = \bar{a}_n {\rm exp}[i(q-k_n)x]$ where 
$\bar{a}_n=\sqrt{1-(q-k_n)^2}$.  The quantity 
\begin{equation}
\hat{a}_n \equiv \sqrt{\langle |a_n(t)|^2\rangle}/\bar{a}_n
\label{eq:distance}
\end{equation} 
describes the 
`distance' from the initial to final $n^{th}$ state and can 
be thought of as an orthogonal coordinate in an 
$n$-dimensional space.  The unit of measure in this 
space is then 
\begin{equation}
d=\sum_{n} \hat{a}_n^2.
\label{eq:metric}
\end{equation} 
If it is assumed that a phase slip has occurred when $d=1$, 
then it is natural to interpret the relative probability of 
an $n^{th}$ order phase slip as
\begin{equation}
P_n = \frac{\hat{a}^2_n}{\sum_{n} \hat{a}_n^2}.
\label{eq:linearpn}
\end{equation}
Eq.~(\ref{eq:linearpn}) provides a qualitative picture of the state
selection process and makes it possible to compare the linear theory
to numerical results.  This will be done in Section IV (in particular,
see Fig.~\ref{fig:pnvsalpha}).   

In addition to the dependence of $P_n$ on $\lambda_n^+$, $P_n$ also 
depends on the noise strength.  While this is not directly visible from 
Eq.~(\ref{eq:linearpn}) it should be noted that the equation $d=1$ 
imposes a $D$ dependence on $\hat{a}_n$ and $P_n$.

Physically, the noise strength depends on
the temperature of the system via the fluctuation-dissipation theorem. 
The intensity of thermal noise increases as $T \rightarrow T_c$ as demonstrated
by Eq.~(\ref{eq:thermnoise}). Thus, close to $T_c$ 
the relative importance of the noise becomes increasingly important, whereas 
away for $T_c$ the driving force is dominant. Since $\alpha$ has no time 
dependence, the expansion of  $\int_0^t dt_1 \lambda^+_n(t_1,\alpha)$
leads to the same result as obtained by Tarlie and Elder\cite{tarlie98b}, i.e., 
in terms of the intrinsic and extrinsic parameters, the instability
of order $n$ becomes active at time 
$\tau_n= \ell (\partial_q \lambda^+_n \omega \ell)^{-1/2}$.

To summarize, the linear analysis shows that the state selection has a subtle 
dependence on both the applied driving force and on the intrinsic properties 
of the system.   It is important to note that this analysis can only be expected 
to give a qualitative description of the process since the analysis 
does not account for competition between the various modes.  These results will 
be compared with numerical simulations of the stochastic time-dependent GL 
equation in Section~\ref{sec:numerical}. 

\section{Numerical results}
\label{sec:numerical}

The parameters that enter the numerical simulations can 
be estimated by considering typical experimental values, such as 
$T_c=3$K, $T=0.93 T_c$, $H_c=300$G, and $\xi(0)=\sqrt{S}=1000${\AA}. 
With these values the intensity of the noise is $D=10^{-3}$, the GL time is 
$\tau_{GL} = 1.4 \times 10^{-11}$ and $\omega \approx {\cal E} /23 \mu \mbox{V}$. 
In the simulations the temperature is fixed and thus the intensity of noise is
fixed. $\omega$ was varied between 0.0001 to 0.1.  This corresponds to
electromotive forces from $2$nV to $2\mu$V.  For dirty superconductors 
the normal state resistivity can vary between $0.01$ and $1.0\mu\Omega -cm$ and 
the $\rho_o$ varies from $1.0$ to $100.0\mu\Omega -cm$. 
Using these values the dimensionless resistivity, 
$\alpha \approx 10^{-4}-1.0$, depending on the dimensions and the material.

A simple Euler algorithm was used for the time integration
of Eq.~(\ref{eq:model}), and Eq.~(\ref{eq:echem}) was solved in Fourier space. 
The complex order parameter was separated into its real and imaginary parts. 
The simulation parameters were: $L=64$, $dx=0.85$, $dt=0.2$,
where $dx$ and $dt$ are the smallest discrete elements of space and
time respectively.

\begin{figure}[!]
\centerline{
\epsfxsize=\columnwidth  
\epsfbox{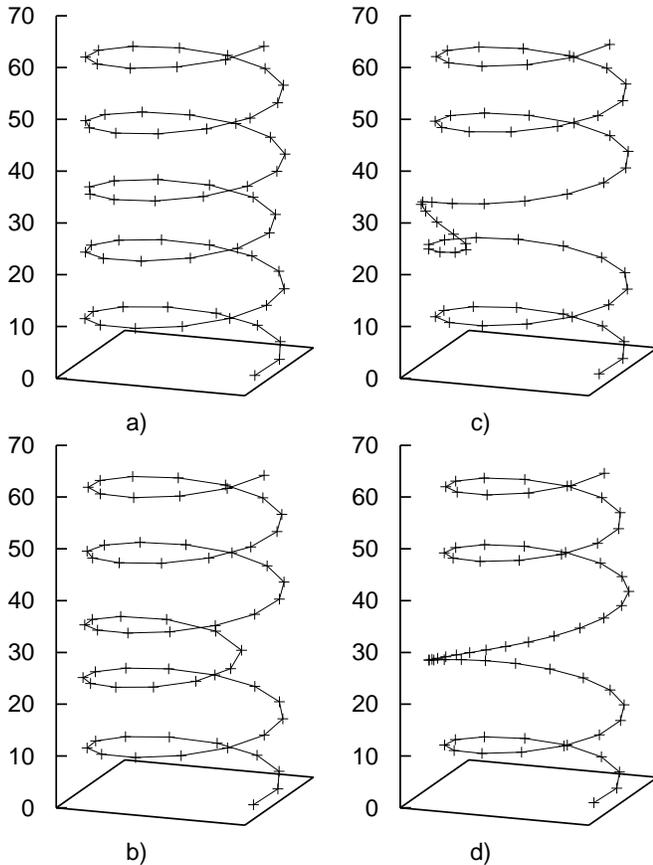}
}
\caption[Snapshot of a phase slip in a superconducting ring]
{A snapshot of a phase slip process with $L=64$ and $n_{\ell}=5$.
a) The current carrying states are uniformly twisted
plane waves. b) Due to fluctuations, the supercurrent at some
location along the wire grows slightly faster than in the
rest of the system. c) At a phase slip the system makes a transition
back to 
below the critical current by reducing the number of loops
in the helix. d) After the phase slip, the system retains the perfectly
superconducting state with $J<J_c$ everywhere.
z-axis: length, x-y plane: Re$[\Psi]$ and Im$[\Psi]$.}
\label{fig:snap}
\end{figure}

A useful parametrization of the length of the system is
$n_{\ell} \equiv \ell q_c /2 \pi$, where $q_c=1/\sqrt{3}$ from the 
Eckhaus analysis of the GL equation as discussed above. $n_{\ell}$ is 
interpreted as the winding number of the order parameter when the 
Eckhaus instability is encountered. For the 
simulations to follow $n_{\ell}=5$ (see Fig.~\ref{fig:snap}). 
This allows enough complexity due to 
interaction between
different modes, i.e. five modes can compete for occupation, 
while remaining numerically tractable. 
When computing the probability of an $n^{th}$ order phase slip, $P_n$, 
the averaging was typically from 
2000 phase slip events (small $\omega$) up to 15,000 phase
slips (large $\omega$).  Simulations performed at large values 
of $\alpha$ and $\omega$ (not shown here) often lead to unusual 
results which may be due to numerical inaccuracies. 

\subsection{Dynamics of the order parameter}

\begin{figure}[!]
\centerline{
\epsfxsize=\columnwidth
\epsfbox{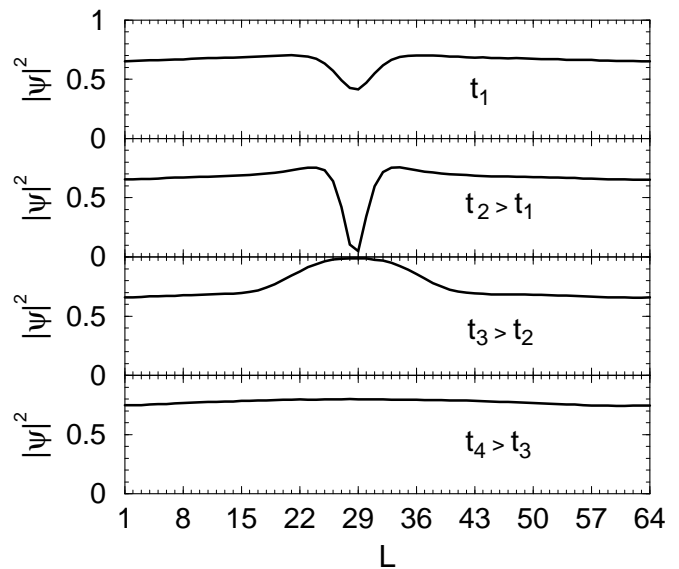}
}
\caption[The behavior of the order parameter at a phase slip]
{Order parameter just before and after the phase slip. As $|\Psi|\rightarrow 0$,
the phase gradient must grow in order to maintain a constant current. As the amplitude 
goes to zero, it can 'slip' by a multiple of $2 \pi$ and relax to a state of lower current.}
\label{fig:orderp}
\end{figure}

The dynamics of the order parameter around a phase slip is illustrated 
in Figs.~\ref{fig:snap} and \ref{fig:orderp}.  Fig.~\ref{fig:snap} 
illustrates that a current carrying state is a uniformly twisted plane 
wave.  As the current increases the helix becomes more tightly
wound.  Due to fluctuations, there will be weak spots where the local
supercurrent reaches the critical current before the rest of the system.
This is the point where the amplitude of the order parameter starts to 
decay rapidly toward zero.  When $|\Psi|^2 \rightarrow 0$, 
the phase slip center momentarily disconnects the phases
to the left and right of it, the helix looses
a loop, and the supercurrent jumps to a lower value. This cycle is 
repeated periodically.

In Figs.~\ref{fig:snap}  and~\ref{fig:orderp} the behavior 
described using linear analysis in the previous section is clearly visible:
as the supercurrent
increases the absolute value of the order parameter, $|\Psi|^2$, decreases
and at the moment of the phase slip approaches zero.  After the phase slip the
order parameter rapidly recovers.  Fig.~\ref{fig:orderp} demonstrates this behavior. 
This allows the  amplitude to relax toward 
equilibrium in the vicinity of the phase slip center (times $t_2$ and $t_3$ in
Fig.~\ref{fig:orderp} and  Fig.~\ref{fig:snap}c,d). After a short time the wire 
obtains a uniform current  ($t_4$ in Fig.~\ref{fig:orderp}).

        To quantify the phase slip events it is useful to consider two quantities:
the spatially averaged supercurrent and the winding number, both at time \( t \). 
The spatially averaged supercurrent is given by  
\begin{equation}
\label{eq:dscurrent}
J_{s}(t)=\frac{1}{2\ell }\int _{0}^{\ell }J_{s}(\Psi )dx.
\end{equation}
The winding number is a measure of 
the total phase change in the system and can be defined as 
\begin{equation}
\label{eq:winding}
W(t)=\frac{1}{2\pi }\int _{0}^{\ell }dx\frac{\partial \left[ \arg \left( \Psi (x,t)\right) \right] }{\partial x}.
\end{equation}
 As described above, the order-parameter can change its winding number by \( 2\pi n \),
where \( n=\pm 1,\pm 2... \) Only changes by an integral multiple of \( 2\pi  \)
are possible in order to preserve the continuity of the order-parameter. This
also implies that at a single (multiple) phase slip, the system removes exactly
one (integral multiple) fluxoid.

\begin{figure}[!]
\centerline{
\epsfxsize=\columnwidth  
\epsfbox{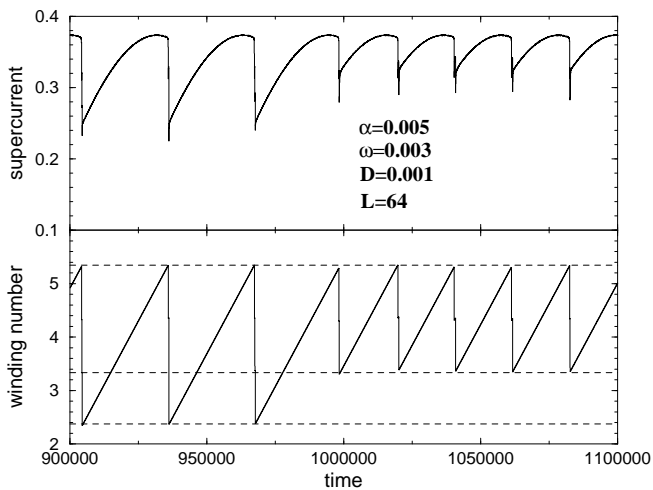}
}
\caption[Supercurrent and winding number as a function of time]
{Supercurrent and winding number as a function of time.
When this figure is compared to the state selection probabilities
in Fig.~(\protect{\ref{fig:selection}}), it can be seen that
the probabilities of double and triple phase slips are almost equal,
but there is still a small probability for single slips. 
}
\label{fig:wind}
\end{figure}

\begin{figure}[!]
\centerline{
\epsfxsize=\columnwidth
\epsfxsize=\columnwidth 
\epsfbox{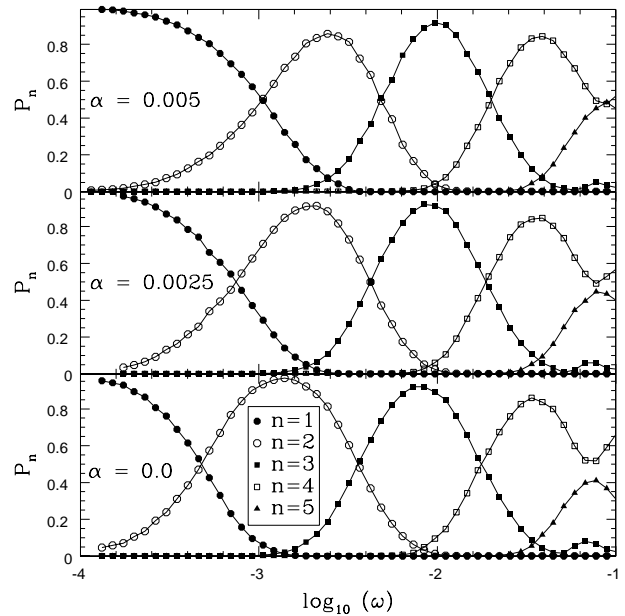}
}
\caption[State selection probability as a function of the driving force]
{State selection probabilities from numerical simulations as a function of 
the driving force. The closed circles denote single phase slips,
open circles doubles, closed squares triples, open squares
quadruples, and closed  triangles phase slips of order five.}
\label{fig:selection}
\end{figure}

Fig.~\ref{fig:wind} displays the time development of the 
supercurrent and winding number defined in 
Eqs.~(\ref{eq:dscurrent}) and (\ref{eq:winding}) respectively.  
The electric field drives the current to the critical current 
where an instability occurs and the current jumps to a 
lower value.  As suggested by  Fig.~\ref{fig:wind},
there can be several modes simultaneously present.
In the figure phase slips of order two dominate but occasionally there
are jumps of order three.  The relative occurrence of phase slips of 
all orders is shown in Fig.~\ref{fig:selection} as a function of 
driving force (i.e., $\omega$) for several values of $\alpha$. 

As discussed in connection with the linear stability analysis
the appearance of phase slips of different order is a subtle issue. 
For example, every now and then the winding 
number displays little dips, as if the total phase slip was a result of a 
two stage process. 
It is instructive to look at the state selection probabilities
in Fig.~\ref{fig:selection} together with the dynamics of the supercurrent and 
the winding number in Fig.~\ref{fig:wind}. 
As seen from  Fig.~\ref{fig:selection} phase slips of order $n=1$ dominate the 
process at low driving forces.  As the driving force 
is increased phase slips of order $n=2$ 
become dominant and the shape of the probability curve of becomes skewed.
Order by order other modes become dominant in a similar manner.
This is consistent with the linear stability analysis as shown 
in Fig.~\ref{fig:linear}.

 \begin{figure}[!]
\centerline{
\epsfxsize=\columnwidth 
\epsfxsize=8cm
\epsfbox{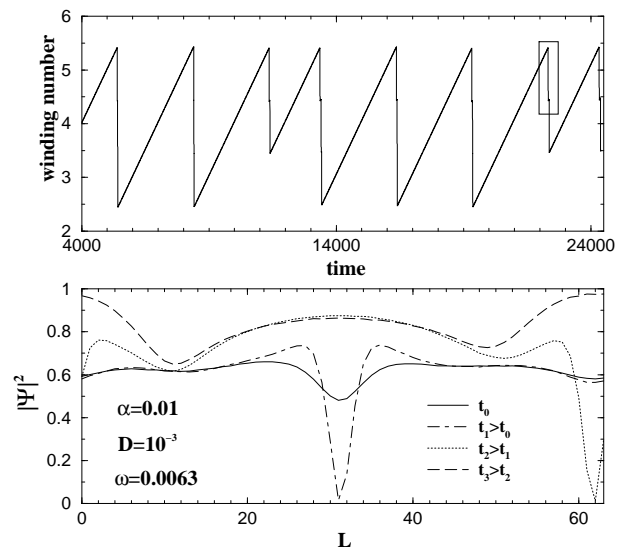} 
}
\caption[Competition between phase slips of different order]
{Upper figure: The winding number as a function of time. 
Lower figure: The square of the amplitude of the order parameter taken at four 
different times inside the boxed area. As the winding numbers increases,
there are spots where the amplitude starts to decay ($t_0$). This leads
to competition and coexistence of several modes, and possible several 
phase slip centers. The parameters correspond to the case when three modes,
$n=1,2,3$, are present.
}
\label{fig:comp}
\end{figure}

The little dips referred to above are a result of competition between the modes. 
As seen in the linear stability analysis, modes
of lower order become unstable first but the higher order ones grow at a faster 
rate. This leads to competition and crossover effects. This implies that 
the dips in  Fig.~\ref{fig:wind} are not due to a result of a two-stage process
where a phase slip of higher order occurs via two lower order processes, but
instead due to the {\it coexistence} of different modes with different growth rates.
Fig.~\ref{fig:comp} illustrates the complicated nature of the phase slip
when several modes are simultaneously present. There is a competition between 
different modes, and it is even possible for several phase slip centers to 
exist (almost) simultaneously.

\begin{figure}[!]
\centerline{
\epsfxsize=\columnwidth
\epsfbox{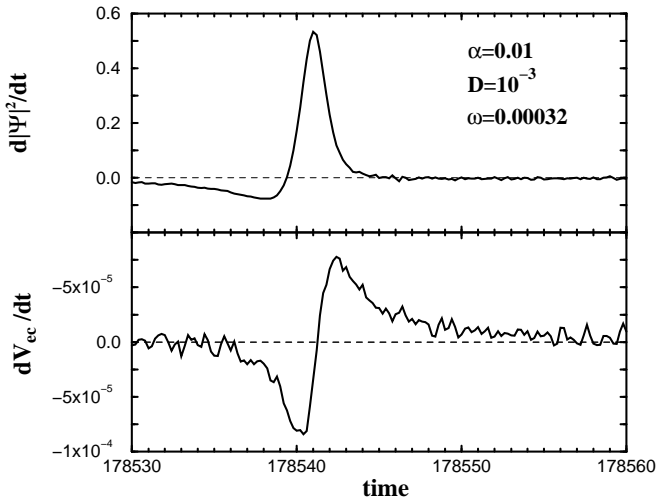}
}
\caption[$\Psi^2$ and the electrochemical potential at phase slip]
{Upper figure: The time derivative of $|\Psi|^2$ at the location of the phase slip as a function of time.
Lower figure:  The rate of change the electrochemical potential at the phase slip center as a function of time.
The simulation parameters correspond to a case when single slips dominate almost completely.
The slice runs from immediately before the phase slip to just after it.}
\label{fig:vec}
\end{figure}

Fig.~\ref{fig:vec} shows the time rate of change of $|\Psi|^2$ and the 
electrochemical potential at 
the phase slip center as a function of time, when the mode $n=1$ is dominant. 
The time frame is selected in such a way that the figures cover the immediate 
vicinity of the phase slip.
At the moment of the phase slip,  $|\Psi|^2 =0$. After the phase slip, $\Psi$
rapidly recovers its equilibrium value. Since there is a constant emf acting on the 
superconductor, $|\Psi|$ starts to decrease after its recovery. As seen in the lower
figure, $v_{ec}$ regains its equilibrium value ($v_{ec}=0$) at the phase slip
center  considerably slower than $\Psi$. This can be seen in the following way. 
The electrochemical potential is zero if the current is uniform throughout the
sample. However, as seen in Figs.~\ref{fig:orderp} and \ref{fig:comp}, the time
required to reach a uniform current is much longer than the time required for healing
of the order parameter at the phase slip center. Physically, this corresponds to 
relaxation of the charge imbalance\cite{btinkham96a} in a superconductor. 
The relaxation is diffusive\cite{btidecks90a,skocpol74a,dolan77a}, 
with time scales typically of order $10^{-9}-10^{-10}$s.

\begin{figure}[!]
\centerline{
\epsfxsize=\columnwidth
\epsfxsize=\columnwidth 
\epsfbox{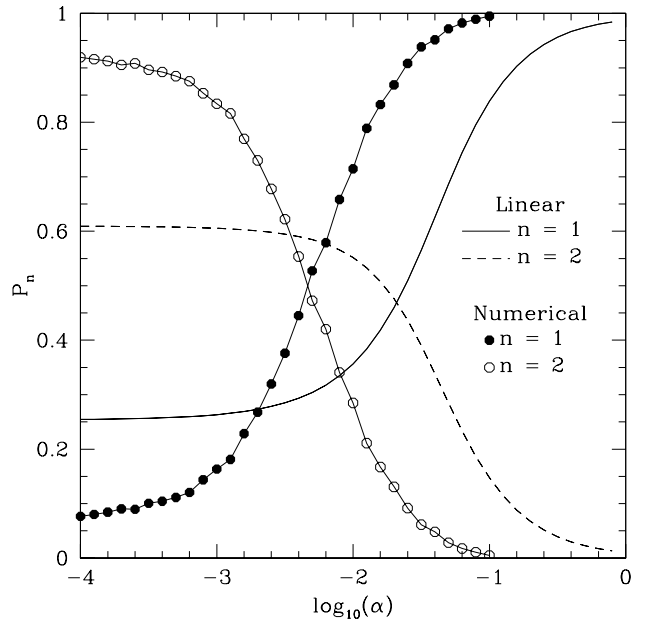}
}
\caption[State selection probability as a function of the alpha]
{State selection probabilities as a function of 
the dissipation (i.e., $\alpha$) for $\omega=10^{-3}$.} 
\label{fig:pnvsalpha}
\end{figure}

        As discussed in the preceding section the electrochemical potential, or 
dissipation, changes the probability of making an $n^{th}$ order phase slip. This can be 
seen in Fig.~\ref{fig:selection} for all values of $\omega$.  To highlight this 
feature the selection probabilities were numerically estimated as a function of 
$\alpha$ for $\omega=10^{-3}$ and are displayed in Fig.~\ref{fig:pnvsalpha}. 
In this figure the linear prediction (i.e., Eq. (\ref{eq:an2}-\ref{eq:linearpn})) 
is also included for comparison. 
While the linear analysis fails to predict
the correct amplitudes for different modes, it provides the correct qualitative picture
and predicts the order in which different modes become dominant. The quantitative 
discrepancies stem from two factors, first the non-linear terms seem to favor a separation 
between the modes. Fig.~\ref{fig:selection} shows that once a mode becomes dominant,
it quickly suppresses all the other ones. However, in linear theory all the allowable
modes have much higher amplitudes at all values of the driving force
$\omega$ \cite{tarlie98a}.
Second, at the limit $\alpha \rightarrow 0$ the linear theory predicts accurately 
the crossover points where a new mode becomes
dominant\cite{tarlie98a}, but the presence of dissipation (finite $\alpha$) has
a significant effect on that as can be seen in  
Figs.~\ref{fig:selection}~and~\ref{fig:pnvsalpha}.

\subsection{Power dissipated at a phase slip}

A phase slip is a dissipative process where electrical energy is locally 
converted into heat.  This is due to Ohmic resistance at 
the phase slip center.
Early experiments\cite{skocpol74a,dolan77a} showed that the differential resistance related 
to the phase slip is temperature independent for a wide range of 
temperatures except very close to $T_c$. This is a delicate issue; heating due
to Ohmic resistance changes the local critical current, and issues related to charge
imbalance and relaxation may become important\cite{btidecks90a}. 
In the following, the heat generated at a phase slip is estimated. 
The normal carriers are assumed to follow  Ohm's law.

The Joule heating law can be used to estimate the heat generated
at a phase slip. The power generated is 
\begin{equation}
P= \int_{volume} \vec{j}_n \cdot \vec{E}\,  d\Omega = S \int j_n E_x dx,
\end{equation}
where $d\Omega$ is a volume element, $S$ is the cross sectional area of the ring,
$\vec{j}_n$ is the dimensional normal current density and  $E_x$ is the electric 
field along the wire.  In terms of the electrochemical potential 
The energy per unit volume can then be written
\begin{equation}
\frac{E}{V} =  {\cal E}_o \int_0^\tau\left[\int_0^{\ell} 
|\vec{J}_n|^2dx\right]d\tau',
\end{equation}
where, ${\cal E}_o = (2H_c^2/l)\alpha$
and $\vec{J}_n$ is the dimensionless normal current density.

\begin{figure}[!]
\centerline{
\epsfxsize=\columnwidth 
\epsfbox{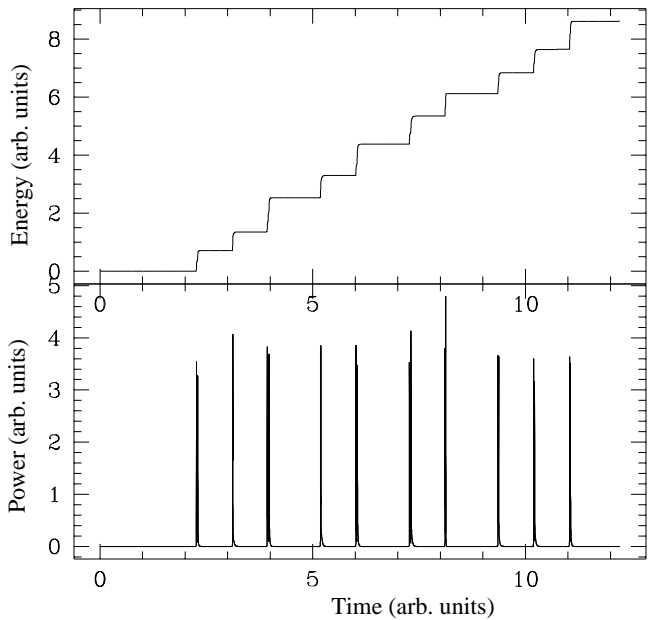}
}
\caption[Power and energy dissipated during a phase slip]
{Upper figure: Energy (in arbitrary units) dissipated as a function of time. 
Lower figure: The corresponding power (in arbitrary units) dissipated at phase slips.
The figures show clearly the quantized nature of dissipation. The crossover
effects are visible in the width of peaks.    
The simulation parameters were $dx=0.85$, $dt=0.2$, $\alpha=0.01$, $\omega=0.0064$
and $D =10^{-3}$. These parameters correspond to a case when three modes ($n=1,2,3$)
are active at the same time.
}
\label{fig:power1}
\end{figure}

The increase in temperature due to a phase slip can be estimated using 
the heat capacity. The heat capacity per unit mass is
$$
c = \frac{1}{m}\frac{\Delta E}{\Delta T} 
$$
where $c$ is the specific heat.  The 
change in temperature is then
\begin{equation}
 \Delta T = T_0 \int_0^\tau \left[\int_0^{\ell}
J_n^2 dx \right] d\tau',
\label{eq:temp}
\end{equation}
where
\begin{eqnarray}
T_0 =  \frac{2 H_c^2}{\ell\,c\,\rho_m } \,  \alpha,\nonumber
\end{eqnarray}
$\rho_m$ is the mass density and $H_c^2 \propto (1-t)^2$, where $t=T/T_c$
\cite{btinkham96a}.
Eq.~(\ref{eq:temp}) can be used to estimate the change in temperature due 
to a phase slip. The linear dependence of $T_0$ on $(1-t)^2$ 
expresses the well known fact\cite{skocpol74a,dolan77a} that close to $T_c$ the effects
of Joule heating are less significant. Evaluation of $\Delta T$ requires information
about the time and the length scales involving $v_{ec}$, and therefore we have 
not estimated it here. Fig.~\ref{fig:power1} shows the accumulated energy and power
dissipated as a function of time.

\section{Conclusion}

Here, the dynamics of accelerated quasi one-dimensional 
superconductors under the influence of a voltage source 
was studied. A constant emf was used to accelerate 
the supercurrent to the critical current, at which point the Eckhaus
instability is encountered and multiple metastable states can compete
for occupation. Each of these competing metastable states corresponds
to a state with a different supercurrent. The transition to a new state of lower 
current involves generation of a resistive phase slip 
center that heals after the phase slip. 
Because the system was driven by a voltage source, it allowed the study 
of very general phenomenon, namely
the relation to the general methods and problems in 
nonlinear dynamics, statistical mechanics and pattern formation.

Linear stability analysis was used to investigate the Eckhaus instability. 
It was found out that within the linear approximation, 
the state selection process is a competition of two factors: the characteristic time
at which a mode $a_n(t)$ becomes unstable, and the growth rates of the other modes. 
For small driving forces, the low-order modes have time to grow and dominate the 
process, whereas for larger driving forces the faster growth rates of high-order 
modes lead to their dominance. In the intermediate region the 
competition leads to crossover. 

Numerical simulations were performed by simulating  
the stochastic time-dependent Ginzburg-Landau equation. 
It was found out that the behavior is consistent with the predictions of the linear
analysis. Although the behavior was qualitatively similar, nonlinearities 
and interaction between the phase slips at higher driving forces and higher
normal current resistivity lead to differences. 

In spite of the simplicity of
the system, it displays rich and complex phenomena, and more analytical and 
numerical studies are needed. To the authors' knowledge, there exists no
systematic method to study state selection in accelerated systems. 
Recent work \cite{mckane01a,tarlie98a}, suggests that the path integral 
method of Onsager and Machlup\cite{onsager53a} may offer a 
framework for a systematic study of the decay of systems from points of instability
when multiple modes compete for occupation. The extension of this work to problems 
where the dynamical system is evolving in time, as is the case here, has not been 
explored. Additionally, future work could explore the two-dimensional case numerically.

\acknowledgements

This work has been supported by the Academy of Finland (MK), 
the Finnish Cultural Foundation (MK),
the Finnish Academy of Science and Letters (MK), 
Research Corporation grant CC4787 (KRE),
NSF-DMR grant 076054 (KRE),
the Natural Sciences and
Engineering Council of Canada (MG), {\it 
le Fonds pour la Formation de
Chercheurs et l'Aide \`a la Recherche du Qu\'ebec\/} (MG).



\end{document}